%% file: icfp_paper.tex
\documentclass[acmsmall]{acmart}
\AtBeginDocument{%
  }

\input{parts/usepackages.tex}

\input{glossary.tex}

\begin{document}

\title{High-Level Synthesis using SDF-AP, Template Haskell, QuasiQuotes, and GADTs to Generate Circuits from Hierarchical Input Specification}


\input{parts/abstract.tex}

\keywords{High-Level Synthesis, Hardware Synthesis, SDF-AP, Template Haskell, QuasiQuotes, GADTs, Hierarchy}

\maketitle

\input{parts/introductionv2.tex}

\input{parts/sdfap.tex}

\input{parts/conf_relation.tex}

\input{parts/parameterized_buffers.tex}

\input{parts/hof_patterns.tex}

\input{parts/time-area_trade-off.tex}

\input{parts/hierarchy_composition.tex}

\input{parts/toolflow.tex}

\input{parts/case_studies.tex}

\input{parts/related_work.tex}

\input{parts/conclusion.tex}





\bibliographystyle{ACM-Reference-Format}
\bibliography{bibliography}

\end{document}

%% file: parts/usepackages.tex
\usepackage{listings}
\usepackage{glossaries}
\usepackage{subcaption} 
\usepackage{wrapfig} 
\usepackage{xcolor}
\usepackage{tcolorbox}

\newtcbox{\inlinecode}{on line, boxrule=0pt, colback=gray!20, colframe=gray!50, sharp corners, size=fbox, boxsep=2pt}

%% file: glossary.tex
\setacronymstyle{long-short}

\newacronym{adt}{ADT}{Algebraic Data Type}
\newacronym{gadt}{GADT}{Generalized Algebraic Data Type}

\newacronym{mcr}{MCR}{Maximum Cycle Ratio}
\newacronym{sps}{SPS}{Strictly Periodic Schedule}

\newacronym{cdfg}{CDFG}{Control Data Flow Graph}
\newacronym{cfg}{CFG}{Control Flow Graph}
\newacronym{dfg}{DFG}{Data Flow Graph}

\newacronym{dsl}{DSL}{Domain-Specific Language}

\newacronym{df}{DF}{Data-Flow}
\newacronym{hsdf}{HSDF}{Homogeneous Synchronous Data-Flow}
\newacronym{sdf}{SDF}{Static/Synchronous Data-Flow}
\newacronym{csdf}{CSDF}{Cyclo-Static Data-Flow}
\newacronym{sdfap}{SDF-AP}{Static Data-Flow with Access Patterns}
\newacronym{sdfasap}{SDF-ASAP}{Static Data-Flow with Actors with Stretchable Access Patterns}
\newacronym{cp}{CP}{consumption pattern}
\newacronym{pp}{PP}{production pattern}

\newacronym{hof}{HoF}{Higher-order function}

\newacronym{rtl}{RTL}{Register-Transfer Level}
\newacronym{dag}{DAG}{Directed Acyclic Graph}

\newacronym{fsm}{FSM}{Finite-State Machine}
\newacronym{fifo}{FIFO}{FIFO}
\newacronym{fpga}{FPGA}{FPGA}

\newacronym{hls}{HLS}{High-Level Synthesis}
\newacronym{hdl}{HDL}{Hardware Description Language}
\newacronym{svd}{SVD}{Singular Value Decomposition}
\newacronym{dsp}{DSP}{Digital Signal Processor}
\newacronym{scop}{SCoP}{Static Control Parts}
\newacronym{pcp}{PCP}{Post Correspondence Problem}
\newacronym{qor}{QoR}{Quality of Results}
\newacronym{fu}{FU}{Functional Unit}

\newacronym{blast}{BlAsT}{Block Assembly Tool}

\newacronym{repl}{REPL}{Read-Evaluate-Print-Loop}
\newacronym{ast}{AST}{Abstract Syntax Tree}
\newacronym{ghc}{GHC}{Glasgow Haskell Compiler}

\newacronym{ssa}{SSA}{Static Single Assignment}

\newacronym{dct}{DCT}{Discrete Cosine Transform}
\newacronym{com}{CoM}{Center of Mass}

\newacronym{alm}{ALM}{Adaptive Logic Module}

%% file: parts/abstract.tex
\begin{abstract}
Field-Programmable Gate Arrays (FPGAs) provide highly parallel and customizable hardware solutions but are traditionally programmed using low-level Hardware Description Languages (HDLs) like VHDL and Verilog.
These languages have a low level of abstraction and require engineers to manage control and scheduling manually.
High-Level Synthesis (HLS) tools attempt to lift this level of abstraction by translating C/C++ code into hardware descriptions, but their reliance on imperative paradigms leads to challenges in deriving parallelism due to pointer aliasing and sequential execution models.

Functional programming, with its inherent purity, immutability, and parallelism, presents a more natural abstraction for FPGA design.
Existing functional hardware description tools such as Clash enable high-level circuit descriptions but lack automated scheduling and control mechanisms.
Prior work by Folmer et al. introduced a framework integrating SDF-AP graphs into Haskell for automatic hardware generation, but it lacked hierarchy and reusability due to its static buffer definitions.

This paper extends that framework by introducing hierarchical pattern specification, enabling structured composition and scalable parallelism.
Our approach allows engineers to define (high-level) patterns that guide both scheduling and control synthesis.
Key contributions include:
(1) automatic hardware generation, where both data and control paths are derived from functional specifications with hierarchical patterns,
(2) parameterized buffers using GADTs, eliminating the need for manual buffer definitions and facilitating component reuse, and
(3) provision of a reference “golden model” that can be simulated in the integrated environment for validation against the synthesized design.

The core focus of this paper is on the methodology.
But we also evaluate our approach against Vitis HLS, comparing both notation and resulting hardware architectures.
Experimental results demonstrate that our method provides greater transparency in resource utilization and scheduling, often outperforming Vitis in both scheduling and predictability.
\end{abstract}

%% file: parts/introductionv2.tex
\section{Introduction}
Field-Programmable Gate Arrays (FPGAs) are hardware platforms that allow engineers to design custom circuits.
Traditional FPGA development relies on low-level hardware description languages like VHDL or Verilog.
These languages describe circuit behavior and structure in detail, and synthesis tools then convert these descriptions into bitstreams, configuration files that program the FPGA hardware.

However, the low abstraction level of VHDL and Verilog poses challenges for developers, especially as designs grow in complexity.
To address this, many have sought higher-level abstractions.
One popular avenue has been to adapt imperative programming paradigms, such as using C/C++ code as input to \gls{hls} tools, aiming to achieve speedup by leveraging the programmer's familiarity with such languages.
Yet, this approach introduces significant challenges, such as accurately deriving data dependencies and parallelism\cite{edwards2006challenges}.
Identifying true data dependencies in languages that support pointers is complicated by the undecidable pointer aliasing problem\cite{landi1992undecidability,ramalingam1994undecidability}.
FPGAs do not follow the sequential execution model of von Neumann architectures.
Instead, they excel at parallelism and pipelining.
For this reason, functional programming, with its emphasis on purity, immutability, and inherent parallelism, offers a better conceptual match\cite{Boutros2021,Sheeran2005}.

Several functional approaches have been explored to bridge the gap between high-level design and FPGA synthesis, including Lava\cite{Bjesse1998}, Bluespec\cite{Nikhil2008}, and Clash\cite{baaij2010clash}.
Clash, in particular, is a functional language that translates Haskell descriptions into VHDL or Verilog, which can then be synthesized for FPGA implementation.
The concepts of \gls{adt}, \gls{hof}, and function composition lift the level of design abstraction\cite{Baaij,Wester2017thesis}.
However, Clash primarily performs a structural translation, it directly converts high-level functional constructs into hardware descriptions without making design decisions about scheduling or control.
Engineers are left to manually design control mechanisms and optimize scheduling, tasks that become increasingly burdensome as circuits scale in size and complexity.

Folmer et al. introduced a framework that integrates a formal model known as \gls{sdfap} and Haskell to automatically generate hardware circuits\cite{Folmer2022high}.
However, the framework has a few key shortcomings, such as the lack of hierarchy and a limited reusability of node definitions due to the static nature of the generated buffers.
To address these limitations, we propose to extend the framework by introducing hierarchy and a new way of specifying patterns.
Our approach enables engineers to specify (high-level) patterns that guide both scheduling and control signal generation.
By incorporating hierarchy into functional specifications, we offer a structured way to manage complexity and enable the reuse of (hierarchical) components without the need for redefinition.

Our main contribution is the introduction of hierarchy into functional hardware specifications, enabling scalable hardware generation with automated scheduling and control.
To achieve this, we present the following key innovations:
\begin{itemize}
\item Hierarchical pattern specification: We introduce an expressive system for specifying hierarchical patterns in input descriptions.
\item Automatic hardware generation: Data and control paths are generated based on functional description and (hierarchical) patterns.
\item Reusing (sub)components: By leveraging \gls{gadt}s to generate parameterized FIFOs, we enable the automatic generation of local buffers. This eliminates the need for manual buffer specification and offers reusability of (sub)components.
\item Simulation and testing framework: Our approach integrates with Clash's interactive environment to support simulation and verification of (sub)components. This ensures correctness by providing a “golden standard” variant of the system, free of buffers, for comparison.
\end{itemize}
The core focus of this work is on the methodology, but we also compare both notation and resulting architecture with hierarchical specifications using the \gls{hls} tool Vitis.
The results demonstrate transparency in both time and resource consumption for our approach compared with Vitis.


%% file: parts/sdfap.tex
\section{SDFAP}
\label{sec:sdfap}
\gls{sdf} graphs are computational models designed for the analysis of a system's temporal behavior\cite{deGroote2016thesis,lee1987df}.
An \gls{sdf} graph is composed of nodes and directed edges.
Each node consumes data (tokens) from each connected edge at a fixed rate once it begins execution (firing).
Upon completing its execution, the node produces data on the output edges at a predetermined rate.
The use of fixed data rates facilitates static analysis and enables the scheduling of tasks within the system.

One limitation of the \gls{sdf} model is the absence of a firing rule that governs the production and consumption of data across consecutive cycles, a feature often required in hardware implementations.
The \gls{sdfap} model addresses this limitation by introducing the concept of \textit{access patterns}\cite{Tripakis2011,Ghosal2012}.
These patterns define the number of tokens produced or consumed during each firing phase of a node.
Additionally, the model enforces a new firing rule: once a node begins execution, it must complete all of its input patterns before finishing the firing phase.
More formally: The \gls{sdfap} model \(M = (N,E)\) consists of a set of nodes \(N\) and a set of edges \(E\).
\(E\) is defined as a set of directed edges \(e=(n_i,n_j,pp,cp)\) where \(n_i\) and \(n_j\) are nodes in \(N\).
\(pp\) and \(cp\) are the production and consumption pattern respectively.
The length of \(pp\) and \(cp\) must equal the number of clock cycles that respectively \(n_i\) and \(n_j\) take to complete one firing.


%% file: parts/conf_relation.tex
\section{Conformance relation Functional language, SDF-AP, and hardware}
\label{sec:conf_relation}
\begin{figure}
  \begin{minipage}{0.6\textwidth}
    \begin{subfigure}[h]{\textwidth}
      \includegraphics[width=\textwidth]{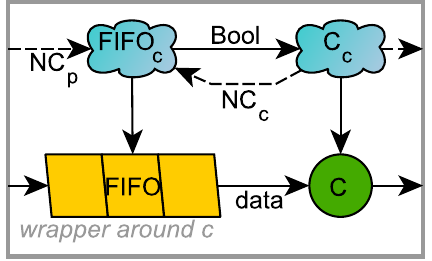}
      \caption{Hardware schematic}
      \label{fig:c_with_wrapper}
    \end{subfigure}
  \end{minipage}
  \begin{minipage}{0.35\textwidth}
    \centering
    \begin{subfigure}[h]{\textwidth}
      \begin{lstlisting}
c ([3],i) = ([2],o)
  where ...
      \end{lstlisting}
      \caption{Annotated code}
      \label{lst:c_with_annotations}
    \end{subfigure}
    \centering
    \begin{subfigure}[h]{\textwidth}
      \includegraphics[width=\textwidth]{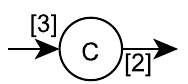}
      \caption{SDF-AP graph}
      \label{fig:c_node_sdfap}
    \end{subfigure}
  \end{minipage}
  \caption{Conformance relation code, graph, and hardware}
  \label{fig:conf_relation}
\end{figure}

To generate hardware based on the functional input description and \gls{sdfap} we have defined a mapping to hardware.
Figure~\ref{fig:conf_relation} depicts the hardware schematic that is generated based on the functional input and the \gls{sdfap} graph.
The combinational hardware described by the function body from Listing~\ref{lst:c_with_annotations} is contained in the green circle $C$.
An engineer has to annotate the definition with patterns in a tuple so that it becomes an \gls{sdfap} node.

\begin{wrapfigure}{R}{0.4\textwidth}
  \centering
  \includegraphics[width=0.4\textwidth]{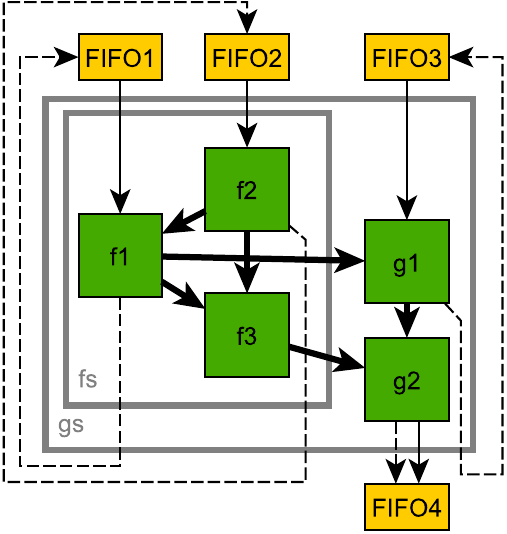}
  \caption{Hierarchical structure of composed hardware.}
  \label{fig:comp_routing_backpressure}
\end{wrapfigure}

Each node is controlled by a local controller $(C_c)$, that manages the execution of the node based on signals from its connected buffers.
The controller receives input signals $(Bool)$ from all associated buffers to determine if the node can begin execution (referred to as "firing").
Once the required conditions are met, the controller initiates the node's operation and manages both its input and output behavior following defined patterns.
Edges in the \gls{sdfap} graph are translated into FIFO buffers in hardware.
Each buffer is managed by a local  $(FIFO_c)$ that ensures compliance with constraints specific to the corresponding edge.
The FIFO controller is responsible for signaling the node controller whether, based on the FIFO content and edge constraints, the node is ready to fire $(Bool)$.
The node controller from the producing node communicates its firing phase to the receiving buffer $(NC_p)$.
Notably, there is no need for the FIFO buffer to send acknowledgment signals back to the producing node controller.
This is because it is assumed that sufficient buffer capacity is always maintained, due to the \gls{sdfap} schedule, eliminating the necessity for backpressure mechanisms.
However, the FIFO controller requires a signal $(NC_c)$ from its consuming node controller $(C_c)$ regarding the current firing state of the consuming node.
For functions with multiple input edges, each edge is assigned a dedicated FIFO buffer and FIFO controller.
All FIFO controllers associated with a node independently signal the single node controller, which uses these inputs to determine the overall readiness of the node to fire.

When composing multiple functions hierarchically, as depicted in Figure~\ref{fig:comp_routing_backpressure}, control signals (dashed lines) to the FIFOs must be routed from the function outputs to the specific FIFOs.
In functional languages like Haskell, functions can have multiple inputs but only a single output.
However, tuples or vectors can be used to bundle multiple output signals.
In Figure~\ref{fig:comp_routing_backpressure}, the \textit{gs} function consists of \textit{g1}, \textit{g2}, and \textit{fs}, while \textit{fs} is internally composed of \textit{f1}, \textit{f2}, and \textit{f3}.
All the black arrows are also FIFOs, and hence also have control signals back and forth, but are left out in the figure.
Two of the three input arguments of \textit{gs} are passed to \textit{fs}, which forwards them to \textit{f1} and \textit{f2}.
These nodes (\textit{f1} and \textit{f2}) must signal back to their respective FIFO controllers regarding their firing states, enabling the FIFO controllers to determine when data has been consumed.
In a hierarchical composition like this, these signals must be routed through multiple levels, from the output of \textit{fs} to the output of \textit{gs} and finally to the specific FIFOs.
As the hierarchy deepens, signal routing can become increasingly complex and messy.
To address this, we incorporate the buffers directly into the function wrappers at the lowest level.
The lowest level in our case means the lowest annotated function.
This local integration of buffers simplifies the routing of control signals and ensures efficient management of hierarchical compositions in hardware designs.

%% file: parts/parameterized_buffers.tex
\section{Introducing hierarchy: Parameterized buffers}
\begin{wrapfigure}{L}{0.4\textwidth}
  \includegraphics[width=0.4\textwidth]{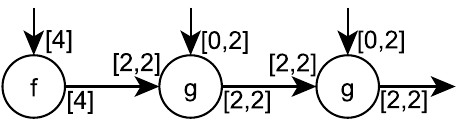}
  \caption{Composition of \textit{f} and \textit{g}}
  \label{fig:comp_diff_buff}
  \vspace{-1em}
\end{wrapfigure}
The main challenge addressed by this paper is managing hierarchy.
As discussed in Section~\ref{sec:conf_relation}, every node is locally defined using controllers on edges and nodes.
This ensures that all back-and-forth signaling between the FIFO buffer and the consuming node occurs within the wrapper.
However, the decision to integrate local buffers within wrappers introduces a new challenge: functions reused in multiple parts of the code may require different buffers.

An example is depicted in Figure~\ref{fig:comp_diff_buff}, where a composition of \textit{f} and two instances of \textit{g} is shown.
As a consequence, the FIFO between \textit{f} and \textit{g} differs from the FIFO between the two instances of \textit{g}.

In our HLS tool, we generate a new function definition with a wrapper around the function that includes FIFOs.
However, we aim to avoid creating distinct wrappers for every instantiation of the same function.
To achieve this, we introduce parameterized buffers within the wrapper, where the final parameters for a FIFO can be set during instantiation.
This allows the FIFO between \textit{f} and \textit{g} to have different parameters than the FIFO between the two instances of \textit{g}.
In the example of Figure~\ref{fig:comp_diff_buff}, the output pattern of \textit{f} is [4], while the input pattern for \textit{g} is [2,2].
This implies that the FIFO between \textit{f} and \textit{g} must convert a vector of length 4 into a vector of length 2.
The vector length indicates the amount of wires synthesized on the FPGA.
In the case of the FIFO between the two instances of \textit{g}, where the output pattern matches the input pattern, no conversion is required.
In hardware, this means that one instance of \textit{g} has as input a vector of length 2, while another instance has as input a vector of length 4.
To support such flexibility in input type, we utilize \gls{gadt}s for buffer instantiation within the wrappers.
\gls{gadt}s enable us to parameterize buffers, specifying properties like the lengths and the data types for inputs and outputs.
This allows us to reuse the same wrapper for every node and determine the specific parameters during function instantiation.

\begin{lstlisting}[label={lst:wrapper_c}, caption={Generated wrapper for \inlinecode{c}}]
c_wrapper (gadt1, inp1) (gadt2, inp2) ...  = (nc',out)
  where
    nc = register Idle nc'
    nc' = updateNc nEn nc
    (bool1,data1) = fifoAndController gadt1 [2] nc' inp1
    (bool2,data2) = fifoAndController gadt2 [4] nc' inp2
    ...
    nEn = bool1 && bool2 && ...
    out = c data1 data2 ...
\end{lstlisting}
Under the hood, the compiler automatically generates a wrapper function for each \gls{sdfap} node, encapsulating its execution logic.
This wrapper function integrates all components depicted in Figure~\ref{fig:c_with_wrapper}, with a corresponding code snippet shown in Listing~\ref{lst:wrapper_c}.
\begin{itemize}
  \item Input Handling (Line 1): The function receives an incoming tuple containing both the \gls{gadt} (which defines parameters that may alter the input type) and the input data consisting of the node status $(NC)$ from the producing node $(NC_p)$ and the actual data.
  \item NC Update (Lines 3-4): The node status $(NC_c)$ is updated based on the enable signals (Boolean values) generated by the FIFO controller.
  \item FIFO Control (Lines 5-6): The function \inlinecode{fifoAndController} is instantiated with the \gls{gadt} configuration, patterns from the \gls{sdfap} graph, node status $(NC)$, and input data. This determines an enable signal (boolX), indicating whether the node can fire according to the FIFO constraints, and produces the corresponding data for the function.
  \item Signal Bundling and Function Execution (Lines 8-9): The enable signals are bundled, and the function is applied to the incoming data, producing the output.
\end{itemize}
Using Template Haskell, the compiler dynamically instantiates the required number of FIFO components (\inlinecode{fifoAndController}) matching the number of inputs.
This also applies to the generated enable signals that are bundled into one \inlinecode{nEn} signal.
Due to this wrapper mechanism, the input type of the wrapper function is different from the original function.
A function originally defined as \inlinecode{c :: a -> b -> c} is now assigned a type that depends on the \gls{gadt}, generally expressed as: \inlinecode{(GADT,(NC,a)) -> (GADT, (NC,b)) -> (NC,c)}
Since both the wrapper and non-wrapper versions of the function coexist, the golden standard (non-wrapper) function can still be tested independently and used as a reference for comparison against the wrapper function, which incorporates all control mechanisms.
This principle of local wrappers allows us to also employ these wrapper functions inside \gls{hof}s, introducing hierarchy at the \gls{hof} level, which is discussed in the next section.

%% file: parts/hof_patterns.tex
\section{Introducing hierarchy: HoF patterns}
To employ \gls{hof}s, the engineer must specify a distinct pattern associated with the \gls{hof}.
In our tool, this pattern is provided as the first argument to the \gls{hof}, as demonstrated in Listing~\ref{lst:hof_patterns}.
\begin{minipage}{0.48\textwidth}
\centering
\begin{lstlisting}[label={lst:hof_patterns}, caption={Annotated \gls{hof}}]
g xs = os where
  os = map [3] f xs
\end{lstlisting}
\end{minipage}
\begin{minipage}{0.48\textwidth}
\centering
\begin{lstlisting}[label={lst:hof_patterns_h}, caption={Hierarchical annotated \gls{hof}s}]
foo xss = oss where
  oss = map [2,2] bar xss
  bar xs = map [1,1,1] f xs
\end{lstlisting}
\end{minipage}

\begin{figure}
  \centering
  \subfloat[\centering{\inlinecode{map [3] f xs}}]{
    \centering
    \includegraphics[height=10em]{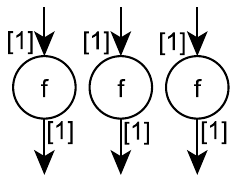}
    \label{fig:map3f}
  }
  \subfloat[\centering{\inlinecode{foldl [3] n s xs}}]{
    \centering
    \includegraphics[height=8em]{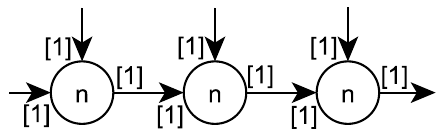}
    \label{fig:foldl3n}
  }
  \caption{\gls{sdfap} graphs of \gls{hof}s}
  \label{fig:sdfap_hofs}
\end{figure}

During code analysis, the our \gls{hls} tool constructs a hierarchical pattern representation.
The definition of a pattern is shown in Listing~\ref{lst:pattern_def}.
Using this recursive pattern definition, a hierarchical notion of depth is introduced.
In our work, we use the $(|)$ notation to show the levels in the hierarchy, for example, $([2,2]|[1,1,1])$, depicts the pattern for the input (\inlinecode{xss}) for the hierarchical \inlinecode{foo} node in Listing~\ref{lst:hof_patterns_h}.
It is important to note that a hierarchy of \gls{hof}s, that construct a \gls{hof} pattern cannot be seen as a single \gls{sdfap} node that adheres to the strict firing rules of \gls{sdfap}.
\begin{lstlisting}[label={lst:pattern_def}, caption={Recursive pattern type}, deletekeywords={Integer}]
data Pattern where
  DefP :: [Integer] -> Pattern
  HierP :: [Integer] -> Pattern -> Pattern
\end{lstlisting}
When an \gls{sdfap} node is inside a \gls{hof} it results in multiple instances of that node in the corresponding \gls{sdfap} graph.
For example, the map in Listing~\ref{lst:hof_patterns} leads to three instances of the \gls{sdfap} node corresponding to \inlinecode{f}.
If \inlinecode{f} is a node that has both input and output pattern $[1]$, then the resulting \gls{sdfap} graph is shown in Figure~\ref{fig:map3f} where the input and output pattern are both $([3]|[1])$.

For \gls{hof}s that have inner dependencies, such as \inlinecode{foldl}, the resulting \gls{sdfap} graph contains additional edges between instances of the function nodes, but only when there is an \gls{sdfap} node inside the \gls{hof}.
If the function inside the \gls{hof} is just combinational logic, then the entire \gls{hof} is treated as a single \gls{sdfap} node.
As shown in Figure~\ref{fig:foldl3n}, \inlinecode{foldl [3] n s xs} produces three \inlinecode{n} nodes arranged sequentially, assuming that \inlinecode{n} is an \gls{sdfap} node, with edges representing dependencies between them.
This means that the execution latency increases proportionally to the number of instances, as it takes three clock cycles for the final result to appear on the output.
In effect, \inlinecode{foldl} serves as an abstract way of introducing pipelining into the circuit.
In our work, we depict \gls{hof}s with a circle around the function inside the \gls{hof}, visually distinguishing their hierarchical nature as shown in Figure~\ref{fig:3d_hofs}.
However, in hardware, this hierarchical representation translates into multiple function instances with parameterized buffers.
The structured nature of \gls{hof}s, in combination with the hierarchical patterns, offers transparency into resource consumption which is discussed in the next section.

%% file: parts/time-area_trade-off.tex
\section{Hierarchy with HoFs, a matter of resource consumption}
\begin{wrapfigure}{L}{0.3\textwidth}
  \includegraphics[width=0.3\textwidth]{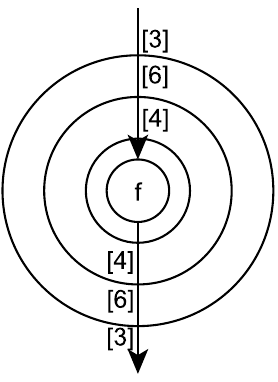}
  \caption{3-dimensional \gls{hof}s with patterns}
  \label{fig:3d_hofs}
\end{wrapfigure}
When utilizing \gls{hof}s within a hierarchical structure, the transparency into its resource consumption is clear.
For instance, consider a function \inlinecode{f} that operates on a three-dimensional vector of size $3 \times 6 \times 4$.
Using \gls{hof}s (multiple \inlinecode{map} nested), as illustrated in Figure~\ref{fig:3d_hofs}, \inlinecode{f} can be applied to every element of this three-dimensional vector.

If \inlinecode{f} represents a square function, each operation requires a single multiplication, which in hardware corresponds to a \gls{dsp} block on an FPGA.
Using our \gls{hls} tool, the engineer can adjust the pattern specified for the \gls{hof} at every level to control the extent of hardware usage and computation time.
For example:
\begin{itemize}
  \item Pattern $([3]|[6]|[4])$ results in all 72 multiplications performed in parallel, requiring 72 \gls{dsp}s. The computation is completed in a single clock cycle.
  \item Pattern $([1,1,1]|[6]|[4])$ requires 24 \gls{dsp}s. However, if the input remains a  $3 \times 6 \times 4$ vector, then the latency increases to 3 clock cycles, as the computations in the top \gls{hof} are serialized.
  \item Pattern $([1,1,1]|[3,3]|[2,2])$ further reduces the resource consumption, reducing the \gls{dsp} count to 6, but increasing the latency to 12 clock cycles.
\end{itemize}
These modifications only require the engineer to update the pattern, while our \gls{hls} tool handles the instantiation of local buffers with the correct parameters.
Due to the transparent structure of \gls{hof}s, it is predictable what the resource consumption would be when modifying the patterns.
Since the compiler preserves the original definition and generates a separate wrapper function, the original version remains available as a reference model.
In Clash's interactive environment, both the unmodified definition and the generated wrapper, complete with control logic, can be tested and verified against each other.

%% file: parts/hierarchy_composition.tex
\section{Hierarchy in composition}
\begin{figure}
  \includegraphics[width=\textwidth]{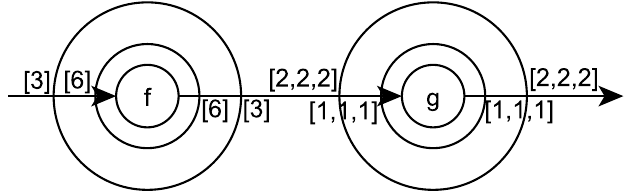}
  \caption{Composition of hierarchical \gls{hof}s}
  \label{fig:comp_hofs}
\end{figure}

Reusing, composing functions, and varying patterns, may lead to different vector sizes as input, and buffers must accommodate these variations.
Consequently, the input type of a function must adapt to the patterns applied.
To reuse the same specification across multiple instances, buffer parameterization is essential.
Consider the code in Listing~\ref{lst:comp_hofs}, Figure~\ref{fig:comp_hofs} is a visual representation of this code.

\begin{lstlisting}[caption={Composition of \gls{hof}s with patterns}, label={lst:comp_hofs}]
os = map [6] (map [3] f) xs
ws = map [2,2,2] (map [1,1,1] g) os
\end{lstlisting}

Here, the output \inlinecode{os}, which has size $6 \times 3$, needs to be reorganized into blocks of length 2 due to the first \gls{hof} in \inlinecode{ws} having the pattern $[2,2,2]$.
Due to the second \gls{hof} having the pattern $[1,1,1]$, these blocks need to be chopped further into blocks of length 1.
The total transformation converts a \inlinecode{Vec 6 (Vec 3 a)} into a \inlinecode{Vec 2 (Vec 1 (Vec 9 a))}.
Each local \inlinecode{g} node (with its wrapper) receives 9 elements into its buffer, computes the \inlinecode{g} function in 9 clock cycles, and outputs a \inlinecode{Vec 2 (Vec 1 a)}.
The wrapper around the \inlinecode{g} function ensures that the buffer is configured to accept 9 elements.

The compilation process handles this as follows:
\begin{itemize}
  \item Instantiate the wrapper function containing the function for both \inlinecode{f} and \inlinecode{g}, equipped with parameterized buffers.
  \item Populate the GADTs with appropriate values, determined by analyzing the hierarchical patterns.
  \item Reshape the input data to align with the hierarchical pattern structure.
\end{itemize}
All these transformations occur in the backend of our tool.
As a result, the engineer only needs to annotate the \gls{hof}s and individual functions with patterns, without worrying about these transformations.
If the specified patterns are incorrect, the engineer is notified for correction.

%% file: parts/toolflow.tex
\section{Toolflow}
\begin{wrapfigure}{R}{0.25\textwidth}
  \begin{lstlisting}[label={lst:quasi_sdfap}, caption={QuasiQuoter}]
[sdfap|
  code here
|]
  \end{lstlisting}
\end{wrapfigure}
To offer the engineer all of these capabilities, we have implemented the following techniques to generate synthesizable Clash code.
Engineers need only surround the portion of their code they want to annotate with patterns using QuasiQuotes as shown in Listing~\ref{lst:quasi_sdfap} and annotate the specification with patterns.
The QuasiQuoter uses Template Haskell to read and modify the \gls{ast} into a description that contains the additional control logic\cite{Mainland2007,sheard2002template}.

\begin{figure}
  \centering
  \subfloat[\centering{AST}]{
    \centering
    \includegraphics[height=10em]{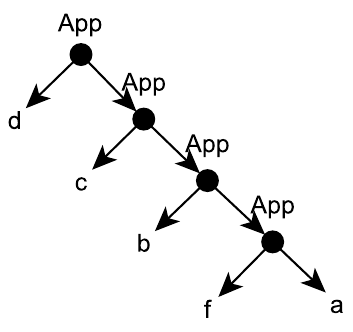}
    \label{fig:expr_ast}
  }
  \subfloat[\centering{DAG}]{
    \centering
    \includegraphics[height=10em]{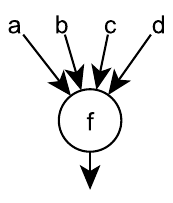}
    \label{fig:graph_f}
  }
  \caption{Expression (\inlinecode{f a b c d}) in both AST and DAG}
  \label{fig:ast_to_dag}
\end{figure}

During the loading of the code in the interactive Clash environment, the following steps occur:
\begin{enumerate}
  \item Code parsing: The code is parsed using both the Clash compiler and the Haskell parser to ensure compatibility and correctness.
  \item Pattern detection: The system detects whether descriptions are annotated with patterns, identifying them as either \gls{sdfap} (hierarchy) nodes or purely combinational logic.
  \item Graph construction: An \gls{sdfap} graph is constructed from the \gls{ast}.
  \begin{itemize}
    \item The AST is parsed and transformed into a \gls{dag}. This transformation extracts a graph as shown in the example in Figure~\ref{fig:ast_to_dag}, where the expression \inlinecode{f a b c d} is shown in both the AST form and \gls{dag}. We traverse the AST using a State Monad to keep track of the nodes and arguments.
    \item Each node is uniquely labeled, as required by the analysis by \gls{sdfap}.
    \item For \gls{hof}s annotated with patterns, a hierarchical pattern (HierP) is introduced.
    \item Patterns from \gls{sdfap} nodes are propagated through all edges that go to non-\gls{sdfap} nodes.
  \end{itemize}
  \item Code transformation: The input description is rewritten into synthesizable Clash code with all the glue-logic and buffers.
  \begin{itemize}
    \item For \gls{sdfap} node declarations, a new function is created with a wrapper containing parameterized buffers (GADTs). The function that is placed inside the wrapper can still be used as a golden standard for verification.
    \item When a node uses an \gls{sdfap} declaration, buffer parameters are determined based on edge patterns from the \gls{sdfap} graph.
    \item Buffer parameters are assigned using lambda functions with partially applied arguments. Each expression containing an \gls{sdfap} node is replaced by a lambda function, where the GADTs are filled with parameters derived from the extracted graph structure.
  \end{itemize}
  \item Simulation and synthesis: The interactive environment in Clash allows simulation of all function definitions, including \gls{sdfap} enhanced versions. Hardware synthesis is performed using the Clash compiler.
\end{enumerate}
Additional advantages:
\begin{itemize}
  \item Subsystem testing: Each subsystem can be tested in an interactive environment.
  \item Golden standard verification: Variants without buffers are still available to test for functional correctness.
  \item Automation: Buffer sizes and boilerplate code are automatically generated, reducing the engineer's workload.
\end{itemize}
This toolflow ensures that engineers can integrate pattern annotations into their hardware designs, enabling synthesis and testing without manual control design.

%% file: parts/case_studies.tex
\section{Case studies}
To evaluate our approach against the current state-of-the-art, we compare it with the Vitis HLS tool, that generates Verilog or VHDL from C code\cite{VitisWeb}.
Engineers using Vitis must annotate their C code with pragmas to guide the synthesis tool.
However, these directives are not always strictly followed by the compiler, potentially leading to unpredictable results.

For a fair comparison, we implemented similar designs in both our approach and Vitis HLS.
Vitis provides latency measurements of the generated circuit, while we obtain resource utilization metrics by synthesizing both the Vitis-generated and our HLS-generated code using Vivado.
We target the Kria FPGA platform, where Vivado synthesizes Verilog into a bitstream that configures the FPGA.
Additionally, Vivado reports the maximum achievable clock frequency, which, combined with latency (measured in clock cycles), determines the total execution time of the circuit.

\subsection{Mapping a square function over a multi-dimensional array}
\label{sec:case_study_maps}
To illustrate hierarchy and expressiveness, we analyze a case study involving the mapping of a square function over a 4-dimensional array.
The full implementation is shown in Listing~\ref{lst:maps_6844}, where each higher-order function (expressed in point-free style for size) is annotated with patterns.
In one case, the specified patterns align with the dimensions of the 4D array, enabling our tool to generate an \gls{sdfap} graph with a latency of just one clock cycle.
Each local square function is assigned a dedicated local buffer, and the Clash compiler automatically translates the specification into Verilog.
For both the \inlinecode{maps6844} and \inlinecode{maps3422} functions, wrappers are generated that can be simulated in Clash's interactive environment.

\begin{minipage}{0.48\textwidth}
  \centering
  \begin{lstlisting}[label={lst:maps_6844}, caption={Maps}]
maps6844 xs = os where
  os = map [6] maps844 xs
  maps844 = map [8] maps44
  maps44 = map [4] maps4
  maps4 = map [4] square
  \end{lstlisting}
  \end{minipage}
  \begin{minipage}{0.48\textwidth}
  \centering
  \begin{lstlisting}[label={lst:maps_3422}, caption={Maps}]
maps3422 xs = os where
  os = map [3,3] maps422 xs
  maps422 = map [4,4] maps22
  maps22 = map [2,2] maps2
  maps2 = map [2,2] square
\end{lstlisting}
\end{minipage}

The synthesis results are presented in Table~\ref{tab:clash_comparison_maps}, where different patterns are examined.
For example, the code in Listing~\ref{lst:maps_3422} is labeled as 3422 and the pattern represents $([3,3] | [4,4] | [2,2] | [2,2])$.
The 1111 label represents a fully sequential computation.
We observe the following resource trends in our \gls{hls}:
\begin{itemize}
  \item As expected, larger patterns result in higher resource utilization (LUTs, registers, and DSPs).
  \item The 1111 pattern leads to full sequential execution, requiring only one DSP and completing the computation in 768 clock cycles.
  \item Conversely, the 6844 pattern enables maximum parallelism, utilizing 768 DSPs to achieve a latency of just one clock cycle.
  \item More parallelism reduces clock cycles but increases routing complexity, lowering the maximum clock frequency.
  \item Despite the clock frequency reduction, the total latency (in nanoseconds) still improves due to the reduced number of cycles.
\end{itemize}

\begin{table}[]
  \begin{tabular}{l||ccc|}
                        & \multicolumn{3}{l|}{Our HLS}  \\ \hline \hline
  Patterns              & 1111     & 3422    & 6844   \\ \hline
  LUTs                  & 62       & 2808    & 13853  \\
  Registers             & 23       & 1350    & 27868  \\
  BlockRAM              & 0        & 0       & 0      \\
  DSPs                  & 1        & 48      & 768    \\
  Clock frequency (MHz) & 353      & 210     & 106    \\
  Latency (cycles)      & 768      & 48      & 1      \\
  Latency ns            & 2178     & 229     & 9
  \end{tabular}
  \caption{Resource consumption for our HLS}
  \label{tab:clash_comparison_maps}
  \vspace{-1em}
\end{table}

For a fair comparison and to assess transparency in Vitis, we implemented two versions of the case study involving the mapping of a square function over a multi-dimensional array.
One case is an implementation using four nested for loops, each annotated with unroll pragmas (pseudo-code shown in Listing~\ref{lst:vitis_mapsLoops}).
The second case is an implementation using separate functions in a hierarchy, each containing a for loop with the same unroll pragmas, partially shown in Listing~\ref{lst:vitis_maps}.
The synthesis results of both implementations are shown in Table~\ref{tab:vitis_comparison_maps}.
We observe that as unrolling increases, there is a moderate rise in LUTs, registers, and sometimes DSPs, but clock frequency remains relatively stable.
Scheduling efficiency in the nested loop variant is significantly better than in the hierarchical version.
Surprisingly, latency in clock cycles fluctuates in the nested loop variant as unrolling increases, whereas it decreases in the hierarchical version.
Despite performing the same computation, Vitis fails to apply the same scheduling strategy for the hierarchical implementation as it does for the nested loop version, resulting in significant differences in latency (ns).


\begin{lstlisting}[label={lst:vitis_mapsLoops}, caption={Maps with nested loops in Vitis}, morekeywords={for,int}]
maps_nested (...)
  loop_inc_a:for (int a = 0; a < Na; a++)
    loop_inc_z:for (int z = 0; z < Nz; z++)
      loop_inc_y:for (int y = 0; y < Ny; y++)
        loop_inc_x:for (int x = 0; x < Nx; x++)
          perform computation.
\end{lstlisting}
\begin{lstlisting}[label={lst:vitis_maps}, caption={Maps with function calls in Vitis},morekeywords={for,int}]
mapsL3 (...)
  loop_inc_z:for (int z = 0; z < Nz; a++)
    call mapsL2

mapsL4 (...)
  loop_inc_a:for (int a = 0; a < Na; a++)
    call mapsL3
\end{lstlisting}

\begin{table}[]
  \begin{tabular}{l||cccc||cccc|}
                        & \multicolumn{4}{l||}{Vitis HLS nested loops} & \multicolumn{4}{l|}{Vitis HLS hierarchical functions} \\ \hline \hline
  Patterns              & 1111     & 3422    & 6844   & 0     & 1111   & 3422  & 6844  & 0     \\ \hline
  LUTs                  & 220      & 365     & 872    & 270   & 211    & 255   & 482   & 262   \\
  Registers             & 280      & 348     & 841    & 249   & 239    & 296   & 418   & 251   \\
  BlockRAM              & 2        & 2       & 2      & 2     & 2      & 2     & 2     & 2     \\
  DSPs                  & 1        & 1       & 4      & 1     & 1      & 2     & 2     & 1     \\
  Clock frequency (MHz) & 271      & 263     & 200    & 222   & 210    & 241   & 215   & 194   \\
  Latency (cycles)      & 772      & 805     & 1076   & 2325  & 5641   & 3685  & 2989  & 2325  \\
  Latency ns            & 2851     & 3057    & 5382   & 3475  & 26902  & 15285 & 13896 & 11983
  \end{tabular}
  \caption{Comparison in resource consumption}
  \label{tab:vitis_comparison_maps}
  \vspace{-1em}
\end{table}

The comparison between our approach and Vitis HLS highlights significant differences in transparency, resource allocation, and scheduling control.
Our pattern-based methodology allows for a clear and predictable scaling of resources, where increasing parallelism leads to an expected rise in the number of allocated components, a corresponding decrease in clock cycles, and an overall improvement in execution time.
In this case study, Vitis HLS does not establish a direct correlation between user-specified pragmas and the generated hardware, making it challenging for engineers to predict resource usage or optimize effectively.
While our approach guarantees that specified patterns are strictly followed, ensuring that the synthesis process adheres to the engineer's intent, Vitis can ignore or inconsistently apply pragmas, leading to unpredictable performance.
This means that using Vitis in this case study, the engineers must resort to extensive manual tuning, including code restructuring and additional annotations, to achieve an optimal design.

\subsection{Center of Mass computation}
\begin{wrapfigure}{R}{0.65\textwidth}
\vspace{-2em}
  \begin{lstlisting}[label={lst:com}, caption={comRows function}, deletekeywords={map,transpose,div}]
coms ims = map [64,64,64,64] com ims

com im = o where
  x = comRows im
  y = comRows (transpose im)

comRows xss = div sumMR sumM where
  m     = map  [8] (fold (+)) xss
  mr    = imap [8] (\i a -> (i+1)*a) m
  sumM  = fold [8] (+) m
  sumMR = fold [8] (+) mr
  \end{lstlisting}
\end{wrapfigure}
To further demonstrate the hierarchical nature of specifications, we present a case study on computing the center of mass for grayscale image blocks.
In this case study, the input consists of 256 image blocks, each of size $8 \times 8$ pixels (Listing~\ref{lst:com}).
The computation is performed using a \gls{hof}, \inlinecode{coms} (Line 1), which applies the \inlinecode{com} function to each block
We use hierarchical annotated \gls{hof}s as shown in Listing~\ref{lst:com}, where the \inlinecode{coms} function maps the \inlinecode{com} function (Line 3).
Here, the function \inlinecode{com} is composed of two parallel applications of \inlinecode{comRows}, which computes the mass distribution across rows.
Since the \inlinecode{fold} operation is not explicitly annotated with a pattern, it is treated as a purely combinational function and is enclosed in a wrapper (Line 8).
Similarly, the lambda function in Line 9 is treated as another combinational function.
The \gls{sdfap} graph gives us the latency of 8 clock cycles when performing a single CoM computation.
However, since each \gls{sdfap} node has its dedicated resources, computations can run in parallel, leading to a pipeline efficiency of 4 clock cycles.
At the top level, \inlinecode{coms} applies \inlinecode{com} 64 times in parallel, processing 64 image blocks in 4 consecutive clock cycles, achieving a 16 cycle latency for the entire batch of 256 images.

Hardware generated from this specification was synthesized, and the resource consumption is shown in Table~\ref{tab:comparison_coms_h8}.
For our HLS, the resource consumption roughly scales in proportion to the size of the patterns.
As the pattern sizes increase, the number of registers, LUTs, and DSPs also increases, while the latency in clock cycles decreases.
This behavior is consistent with the observations in Section~\ref{sec:case_study_maps}, where we analyzed the mapping of a square function.
We observe that the latency in nanoseconds decreases, as the combinational path length remains roughly the same, ensuring that the clock frequency stays constant.
\begin{table}[]
  \begin{tabular}{l||cccc||ccccc|}
                  & \multicolumn{4}{l||}{Our HLS}     & \multicolumn{5}{l|}{Vitis HLS}            \\ \hline \hline
  Patterns        & [1..] & [8..] & [16..] & [64..] & 1      & 8      & 16     & 64     & 0     \\ \hline
  LUTs            & 1316  & 10959 & 21906  & 86962  & 1413   & 2778   & 2798   & 2890   & 1997  \\
  Registers       & 404   & 3080  & 6204   & 24529  & 698    & 1333   & 1348   & 1402   & 1925  \\
  BlockRAM        & 0     & 0     & 0      &  0     & 83     & 100    & 100    & 100    & 66    \\
  DSPs            & 12    & 96    & 192    & 768    & 3      & 6      & 6      & 6      & 12    \\
  Clk freq. (MHz) & 48    & 47    & 46     & 44     & 135    & 140    & 140    & 142    & 146   \\
  Latency cycles  & 260   & 36    & 64     & 16     & 42506  & 21450  & 21418  & 21394  & 796   \\
  Latency ns      & 5367  & 769   & 438    & 181    & 315962 & 153153 & 153460 & 150529 & 5463
  \end{tabular}
  \caption{Comparison in resource consumption}
  \label{tab:comparison_coms_h8}
  \vspace{-1em}
\end{table}
In contrast, for Vitis HLS, the number of DSPs and LUTs remains constant regardless of the applied patterns.
Surprisingly, the version without unrolling pragmas achieves far better performance than the versions with unrolling.
We were unable to determine why Vitis does not utilize additional DSPs when unrolling is applied.
Additionally, the latency and clock frequency results appear extreme and do not correlate with the unrolling pragmas, suggesting that Vitis struggles to determine an optimal schedule in these cases.

%% file: parts/related_work.tex
\section{Related work}
\subsection{FPGA languages}
Sozzo provides an extensive survey of FPGA design languages and tools, categorizing research efforts into \gls{hdl}, \gls{hls} tools, and \gls{dsl}\cite{Sozzo2022}.
The study also includes timelines marking the inception of various tools and their current activity status.


Several tools have explored functional programming approaches for hardware design, including Bluespec, Lava, and Chisel.
Bluespec incorporates a rule-based system with Guarded Atomic Actions, offering a high-level abstraction for hardware synthesis\cite{Nikhil2008}.
Lava, a DSL implemented in Haskell, leverages Haskell's functional programming features to describe hardware circuits declaratively\cite{Bjesse1998}.
Chisel, an HDL embedded in Scala, combines imperative and functional concepts to enable hardware synthesis and includes a C++ simulator for debugging\cite{Bachrach2012}.


Researchers have proposed a distributed memory architecture for dedicated hardware synthesized directly from Erlang programs\cite{Azuma2017}.
This approach generates Verilog HDL from simple Erlang specifications, enabling system construction entirely from functional descriptions.
Additionally, ACAP has been used to produce hardware-software co-design solutions from Erlang, emphasizing the potential for integration between software and hardware\cite{Ishiura2014}.



Aronsson presents a library in Haskell for programming FPGAs, including hardware-software co-design\cite{Aronsson2017}.
Code for software (C) and hardware (VHDL) is generated from a single program, along with the code to support communication between hardware and software.


\subsection{Dataflow formalisms and High-Level Synthesis}
Temporal behavior analysis for hardware design often relies on formal models like \gls{sdf}, introduced by Lee and Messerschmitt\cite{lee1987df}.
In the SDF model, computations are represented as nodes connected by edges, with tokens flowing along these edges.
Each edge is annotated with production and consumption rates, specifying the number of tokens generated or consumed.
A node can "fire" when sufficient tokens are available on all its input edges, and it produces tokens on its output edges according to the defined rates.
This deterministic firing mechanism ensures predictable execution, making SDF a foundational model for hardware design analysis.


The \gls{sdfap} model, employed in this work (see Section~\ref{sec:sdfap}), builds upon \gls{sdf} by introducing access patterns and additional firing rules.
However, \gls{sdfap} has limitations, as identified by Du, who proposes \gls{sdfasap}, an extension that incorporates stretchable access patterns\cite{Du2018, Du2019, Du2020}.
These patterns define an upper bound on data consumption time, allowing for a more flexible interpretation of firing rules.
While \gls{sdfasap} is a promising approach, its stretchability is unnecessary for the strict firing requirements of our framework, but it is extremely interesting candidate for analysis of the hierarchial patterns, this remains future work.
The Block Assembly Tool (BlAsT) applies the principles of \gls{sdfasap} in a visual environment inspired by Simulink, where blocks adhere to SDF-ASAP firing rules, and VHDL code is generated automatically\cite{Du2022}.



Buffer sizing, an essential aspect of dataflow models, is addressed in the work of Honorat\cite{Honorat2024}.
The authors propose two complementary methods for buffer size determination:
First, calculate the theoretical worst-case bounds.
This method tends to overestimate buffer sizes, therefore in the second step, the bounds are refined based on iterative co-simulations.


Dependency issues that cannot be resolved at design time have also led to the introduction of dynamic scheduling supported by dataflow models.
A methodology for automatically generating circuits from C/C++ code integrates dynamic scheduling into hardware synthesis, addressing the unpredictable dependencies encountered in complex applications\cite{Josipovi2022}.


Sinha introduces SynDFG, a generator for scalable Dataflow Graphs (DFGs)\cite{Sinha2015}.
These DFGs are directed acyclic graphs (DAGs) enhanced with steps that emulate both control flow and dataflow, positioning SynDFG as a valuable tool for High-Level Synthesis (HLS) research.


The combination of static and dynamic scheduling has been explored to address the scheduling problem. Static scheduling is applied to well-defined components, which are treated as black boxes\cite{Cheng2020}.
These black boxes are then integrated into a dynamically scheduled dataflow circuit, allowing for a flexible, hybrid approach that combines the predictability of static scheduling with the adaptability of dynamic scheduling.




%% file: parts/conclusion.tex
\section{Conclusion}
This work demonstrates how raising the level of abstraction in hardware design through functional languages can lead to improved transparency, composability, and efficiency.
\gls{hdl}s often function as description tools with significant limitations in hierarchical abstraction, reusability, and mainly scheduling.
Prior work by Folmer et al. introduced temporal analysis using \gls{sdfap} graphs, yet it lacked a structured way to express hierarchical composition and reusable components.

We address these shortcomings by introducing hierarchical pattern specifications that employ parameterized buffers using \gls{gadt}s into our HLS framework.
Our method, using QuasiQuotes, extracts \gls{sdfap} graphs from the specification, determining buffer and scheduling constraints without requiring manual intervention from the engineer.
Patterns can be annotated at both the node definition level and within higher-order functions, providing an explicit and predictable way to control parallelism in hardware generation.
Engineers can reuse (sub)components in separate instances with different GADT parameters, these instances can all be tested and simulated inside the interactive environment of Clash.

Through case studies, we illustrate the transparency of resource allocation when using hierarchical patterns in our framework.
Latency, DSPs, LUTs, and registers all scale predictably with pattern sizes.
Vitis HLS is used where unroll pragmas fail to correlate with resource usage and performance improvements.
Our method not only provides engineers with clear insights into parallelism and resource trade-offs but also outperforms Vitis HLS in the case studies in terms of predictability and scheduling efficiency.